\documentclass[conference,tbtags,10pt,romanappendices]{IEEEtran}

\usepackage[IEEE]{enderheader}
\usepackage[GMACWT,WT]{enderspecheader}

\begin{document}
\thispagestyle{headings}
\title{Achievable Rates for Two-Way Wire-Tap Channels}
\author{
\authorblockN{Ender~Tekin}
\authorblockA{
Wireless Communications and Networking Laboratory\\
Electrical Engineering Department \\
The Pennsylvania State University  \\
University Park, PA 16802\\
tekin@psu.edu \vspace{-.3in}}
\and
\authorblockN{Aylin~Yener}
\authorblockA{
Wireless Communications and Networking Laboratory\\
Electrical Engineering Department \\
The Pennsylvania State University  \\
University Park, PA 16802\\
yener@ee.psu.edu \vspace{-.3in}}
}
\maketitle 

\begin{abstract}
We consider two-way wire-tap channels, where two users are communicating with each other in the presence of an eavesdropper, who has access to the communications through a multiple-access channel.  We find achievable rates for two different scenarios, the \ital{Gaussian two-way wire-tap channel}, (GTW-WT), and the \ital{binary additive two-way wire-tap channel}, (BATW-WT).  It is shown that the two-way channels inherently provide a unique advantage for wire-tapped scenarios, as the users know their own transmitted signals and in effect help encrypt the other user's messages, similar to a one-time pad.  We compare the achievable rates to that of the Gaussian multiple-access wire-tap channel (GMAC-WT) to  illustrate this advantage.
\end{abstract}

\section{Introduction}
Information theoretic secrecy was first developed by Shannon in \cite{shannon:secrecy}. In this work, Shannon showed that to achieve \ital{perfect secrecy} in communications, which is equivalent to providing no information to an enemy cryptanalyst, the \ital{a posteriori} probability of a message must be equivalent to its \ital{a priori} probability.

In \cite{wyner:wiretap}, Wyner applied this concept to the discrete memoryless channel by defining the \ital{wire-tap channel}, where there is a wire-tapper who has access to a degraded version of the intended receiver's signal.  Using the normalized conditional entropy of the transmitted message given the received signal at the wire-tapper as the secrecy measure, he found the region of all possible rate/equivocation pairs, and the existence of a \ital{secrecy capacity}, $C_s$, the rate up to which it is possible to transmit zero information to the wire-tapper.

Reference \cite{leung-hellman:gaussianwiretap} extended Wyner's results to Gaussian channels.  Csisz\'ar and K\"orner, \cite{csiszar-korner:confbroadcast}, improved Wyner's results to weaker, ``less noisy" and ``more capable" channels. Furthermore, they examined sending common information to both the receiver and the wire-tapper, while maintaining the secrecy of private information that is communicated to the receiver only.

In \cite{maurer:secretkeypublicdiscussion}, it is shown that the existence of a ``public" feedback channel can enable the two parties to be able to generate a secret key even when the wire-tap capacity is zero.  More recently, the notion of the wire-tap channel has been extended to parallel channels, \cite{yamamoto:secretsharing}, relay channels, \cite{oohama:relaywiretap}, and fading channels, \cite{barros:fadingwiretap}. Broadcast and interference channels with confidential messages are considered in \cite{liuetal:IBCconf}.  References \cite{liang:genMACconfPAP, liuetal:MACconf} examine the multiple access channel with confidential messages where two transmitters try to keep their messages secret from each other while communicating with a common receiver.  Gaussian multiple-access wire-tap (GMAC-WT) channels are considered in \cite{tekin:ASILOMAR05, tekin:ISIT06, tekin:IT06a, tekin:ALLERTON06}, where transmitters communicate with an intended receiver in the presence of an external wire-tapper.  In \cite{tekin:ISIT06, tekin:IT06a}, we considered the case where the wire-tapper gets a degraded version of the signal at the legitimate receiver, and found the secrecy-sum capacity for the \ital{collective} set of constraints using Gaussian codebooks and stochastic encoders.  In \cite{tekin:ALLERTON06}, the general (non-degraded) GMAC-WT was considered, and an achievable rate region for perfect secrecy with collective secrecy measures was found.

In this paper, we consider the two-way channel where two nodes communicate with each other,  \cite{shannon:twoway}.  We introduce the two-way wire-tap (TW-WT) channel where an external \ital{eavesdropper} receives the transmitters' signals through a general MAC. In particular, we consider the Gaussian Two-Way Wire-Tap Channel (GTW-WT), and the Binary Additive Two-Way Wire-Tap Channel (BATW-WT).  We utilize as our secrecy constraint, the normalized conditional entropy of the transmitted secret messages given the eavesdropper's signal, as in \cite{wyner:wiretap}. We show that satisfying this constraint implies the secrecy of the messages for both users.  In both scenarios, transmitters are assumed to have one secret and one open message to transmit.  We find an achievable \ital{secure rate region}, for both cases, where users can communicate with arbitrarily small probability of error with the intended receiver under \ital{perfect secrecy} from the eavesdropper.

We also show that in cases where a user is not able to achieve secrecy, that user may help the other user increase its secrecy rate or achieve secrecy if it was not possible before, by jamming the eavesdropper.  Thus, similar to the Gaussian multiple-access wire-tap channel, \cite{tekin:ALLERTON06}, \ital{cooperative jamming} helps increase the secrecy rate.
\newlength{\figsize}
\setlength{\figsize}{3in}
\newcommand{\NWt}{\tilde \Nm_\Wch}
\newcommand{\Em}{\v{E}}
\newcommand{\EW}{\Em_\Wch}
\newcommand{\ep}{\varepsilon}
\newcommand{\epW}{\ep_\Wch}
\newcommand{\xor}{\oplus}

\renewcommand{\Wsec}{W}
\renewcommand{\Wmsec}{\Wm}
\renewcommand{\CW}{C_\Wch}

\section{System Model and Problem Statement}
\label{sec:system}
We consider two users communicating in the presence of an intelligent and informed eavesdropper.  Each transmitter $k \in \Ks \triangleq \{1,2\}$ has a secret message, $W_k$, from a set of equally likely messages $\Ws_k=\{1, \dotsc, M_k\}$.  The messages are encoded using $(2^{nR_k},n)$ codes into $\{\tilde X_k^n(W_k)\}$, where $R_k=\ninv \log_2 M_k$.
The encoded messages $\{\tilde \Xm_k\}=\{\tilde X_k^n\}$ are then transmitted.  
Each receiver $k=1,2$ gets $\Ym_k=Y_k^n$ and the eavesdropper $\Zm=Z^n$.  Receiver $k$ decodes $\Ym_k$ to get an estimate of the transmitted message of the other user.  The users would like to communicate with arbitrarily low probability of error, while maintaining perfect secrecy of the messages, $\Wm$.  We assume the channel parameters are universally known, including at the eavesdropper, and that the eavesdropper also knows the codebooks and coding scheme.  We first define \ital{achievability} for this wire-tap channel:
\begin{definition}[Achievable secrecy rates]
\label{def:achrate}
The rate pair $(R_1,R_2)$ is said to be \ital{achievable} for the TW-WT, if for $\e>0$ there exists a code of sufficient length \n such that
\begin{subequations}
\begin{align}
\ninv \log_2 M_k &\ge R_k - \e \quad k=1,2\\
\Perr &\le \e\\
\frac{H(\Wmsec|\Zm)}{H(\Wmsec)} &\ge 1-\e
\end{align}
\label{eqn:achdef}
\vspace{-.08in} 
\end{subequations}
where \vspace{-.02in}
\begin{equation}
\Perr = \frac{1}{M_1 M_2} \sum_{\Wm \in \; \Ws_1\times \Ws_2}
	\prob{\Wmh \neq \Wm |\Wm \text{ was sent}}.
\end{equation}
is the average probability of error for a given code.
\end{definition}

\subsection{The Gaussian Two-Way Wire-Tap Channel}
\begin{figure}[t]
\begin{center}
\resizebox{\figsize}{!}{\input{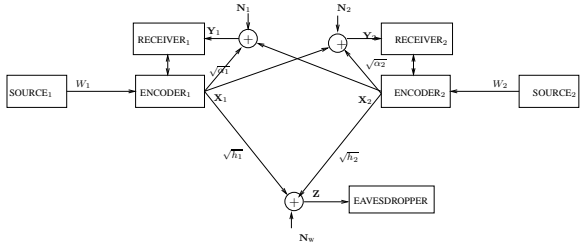}}
\caption{The standardized GTW-WT system model}
\label{fig:GTW}
\end{center}
\vspace{-.3in}
\end{figure}

We describe the Gaussian Two-Way Wire-Tap Channel (GTW-WT), which corresponds to a two-way wireless communications system. We assume slow fading, such that each codeword experiences the same channel coefficient, and also that all parties know the channel coefficients.  The signals at the intended receiver and the eavesdropper are given by
\begin{subequations}
\begin{align}
\Ym_1 &= \tilde \Xm_1 + \sqrt{\hM_2} \tilde \Xm_2 + \tilde \Nm_1 \\
\Ym_2 &= \sqrt{\hM_1} \tilde \Xm_1 + \tilde \Xm_2 + \tilde \Nm_2 \\
\Zm &= \sqrt{\hW_1} \tilde \Xm_1 + \sqrt{\hW_2} \tilde \Xm_2 + \NWt
\end{align}
\label{eqn:GTW}
\end{subequations}
such that $\ninv \sumton{\tilde X_{ki}^2} \le \tilde{\Pmax}_k,$ for $k=1,2$ and $\tilde \Nm_k \isnormal{0,\nvar_k}$ and $\NWt \isnormal{0,\nvar_\Wch}$.
For simplicity, without loss of generality, we consider an equivalent standard form as in \cite{tekin:IT06a} as illustrated in Figure \ref{fig:GTW}.
\begin{subequations}
\begin{align}
\Ym_1 &= \sqrt{\alpha_1} \Xm_1 +\Xm_2 + \Nm_1 \\
\Ym_2 &= \Xm_1 +\sqrt{\alpha_2} \Xm_2 + \Nm_2 \\
\Zm &= \sqrt{h_1} \Xm_1 + \sqrt{h_2} \Xm_2 + \NW
\end{align}
\label{eqn:GTWstd}
\end{subequations}
where, for $k=1,2$, 
\begin{itemize}
\item the codewords $\{\tilde \Xm\}$ are scaled to get $\Xm_1 = \sqrt{\frac{\hM_1}{\nvar_2}} \tilde \Xm_1$ and $\Xm_2 = \sqrt{\frac{\hM_2}{\nvar_1}} \tilde \Xm_2$;
\item the maximum powers are scaled to get $\Pmax_1 = \frac{\hM_1}{\nvar_2} \tilde \Pmax_1$ and $\Pmax_2 = \frac{\hM_2}{\nvar_1} \tilde \Pmax_2$;
\item the transmitters' new channel gains are given by $\alpha_1 = \frac{\nvar_2}{\hM_1 \nvar_1}$ and $\alpha_2 = \frac{\nvar_1}{\hM_2 \nvar_2}$;
\item the wiretapper's new channel gains are given by $h_1 = \frac{\hW_1 \nvar_2}{\hM_1 \nvar_\Wch}$ and $h_2 = \frac{\hW_2 \nvar_1}{\hM_2 \nvar_\Wch}$;
\item the noises are normalized by $\Nm_k = \frac{\tilde \Nm_k}{\nvar_k}$ and $\NW = \frac{\NWt}{\nvar_\Wch}$.
\end{itemize}

\subsection{The Binary Additive Two-Way Wire-Tap Channel}
\begin{figure}[t]
\begin{center}
\resizebox{\figsize}{!}{\input{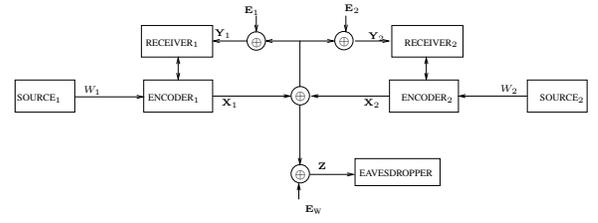}}
\caption{BATW-WT system model}
\label{fig:BATW}
\end{center}
\vspace{-.3in}
\end{figure}
\begin{subequations}
This model, shown in Figure \ref{fig:BATW}, corresponds to a more classical wire-tapped channel, where the binary signals of two transmitters are superimposed on a common wire, as in \cite{shannon:twoway}, and random bit errors are produced as in a binary symmetric channel.  For the BATW-WT, the received signals are given by
\begin{align}
\Ym_1 &= \Xm_1 \xor \Xm_2 \xor \Em_1 \\
\Ym_2 &= \Xm_1 \xor \Xm_2 \xor \Em_2 \\
\Zm_1 &= \Xm_1 \xor \Xm_2 \xor \EW
\end{align}
\label{eqn:BATW}
\end{subequations}
where $\Em_1,\Em_2,\EW$ are n-vectors of binary random variables representing errors such that $\prob{E_{ki}=1}=\ep_k<\onehalf$ and $\prob{E_{\Wch i}=1}=\ep_\Wch<\onehalf$;$\ep_k$ is the error probability at receiver $k=1,2$; $\epW$ is the error probability at the wiretapper.

\newcommand{\Rs}{\s{R}}
\newcommand{\RsGTW}{\Rs^{\scriptscriptstyle{\text{GTW}}}}
\newcommand{\RsBATW}{\Rs^{\scriptscriptstyle{\text{BATW}}}}
\section{Achievable Rates}
In this section, we give our results on some achievable rates for the TW-WT channels considered in this paper.  The proofs for both the GTW-WT and BATW-WT are similar and are summarized in Appendix \ref{app:TWprf}.  For details, please see \cite{tekin:IT07a}.  We first define a few quantities: \vspace{-.05in}
\begin{gather*}
\mmax{\xi} \triangleq \Max{\xi,0}, \qquad
g(\xi) \triangleq \onehalf \log (1+\xi) \\
h(\xi) \triangleq -\xi \log \xi - (1-\xi) \log (1-\xi), \quad 0 \le \xi \le 1\\
\Ps \triangleq \braces{(P_1,P_2) \colon 0 \le P_1 \le \Pmax_1,\, 0 \le P_2 \le \Pmax_2}
\end{gather*}
and
\begin{align}
C_k &= \case{g\paren{P_k}}{\text{GTW-WT}}{1-h(\ep_k)}{\text{BATW-WT}}\\
\CW &= \case{g\paren{h_1P_1+h_2P_2}}{\text{GTW-WT}}{1-h(\epW)}{\text{BATW-WT}}
\end{align}

We now give achievable secret-rate regions for the two channels.  In both channels under consideration, capacity without secrecy constraints can be achieved using independent inputs: for the GTW-WT, this was shown in \cite{han:twoway}. For the BATW-WT, it is easily checked that the symmetry conditions in \cite{shannon:twoway} apply, and the capacity region is a rectangle obtained by equiprobable inputs.  The achievable secrecy rate regions in this paper are obtained using independent channel inputs.
\begin{theorem}
\label{thm:GTWach}
Let 
\begin{align}
\RsGTW(P_1,P_2)= \lbrace (R_1,R_2) \colon \hspace{-1.5in}& \notag \\
	& \begin{array}{l}
	R_k \le g(P_k) \quad k=1,2 \\
	R_1+R_2 \le \mmax{g(P_1)+g(P_2)-g(h_1 P_1+h_2 P_2)} \rbrace
	\end{array}
\label{eqn:GTWachP}
\end{align}
The rate region given below is achievable for the GTW-WT:

\begin{equation}
\RsGTW = \text{convex closure of } \bigcup_{\Pm \in \Ps} \RsGTW(\Pm)
\label{eqn:GTWach}
\end{equation}
\end{theorem} \vspace{-.1in}
\begin{proof} See Appendix \ref{app:TWprf}.
\end{proof}

\begin{theorem}
For the BATW-WT, we can achieve the following set of rates:
\begin{align}
\RsBATW= \lbrace (R_1,R_2) \colon \hspace{-.7in}& \notag \\
	& \begin{array}{l}
	R_k \le 1-h(\ep_k) \quad k=1,2 \\
	R_1+R_2 \le \mmax{1+h(\epW)-h(\ep_1)-h(\ep_2)} \rbrace
	\end{array}
\label{eqn:BATWachP}
\end{align}
\end{theorem} \vspace{-.05in}
\begin{proof} See Appendix \ref{app:TWprf}.
\end{proof}

\section{Maximization of Sum Rate for GTW-WT}
\label{sec:summax}
The achievable regions given in Theorem \ref{thm:GTWach} depends on the transmit powers.  We are naturally interested in the power allocation $\Pmopt=(\Popt_1,\Popt_2)$ that would maximize the total secrecy sum-rate.  Without loss of generality, we will assume that $h_1 \le h_2$.  We formally state the problem as:
\begin{align}
\max_{\Pm \in \Ps} \; C_1+C_2-\CW \notag \hspace{-1in}&\\
	&= \max_{\Pm \in \Ps} \; g\paren{P_1}+g\paren{P_2} - g\paren{h_1 P_1+h_2 P_2} \\
	&\equiv \min_{\Pm \in \Ps} \; \rho(\Pm) 
	\label{eqn:GTWsumprob1}
\end{align}
where
\begin{equation}
\label{eqn:rhodef}
\rho(\Pm) \triangleq \frac{1+h_1 P_1+h_2 P_2}{(1+P_1)(1+P_2)}
\end{equation}
The optimum power allocation is stated below:
\begin{theorem}
\label{thm:GTWsum}
The secrecy sum-rate maximizing power allocation for the GTW-WT is given by:
\begin{equation}
(\Popt_1,\Popt_2) = 
	\begin{cases}
	(\Pmax_1,\Pmax_2), &\text{if } h_1 \le 1+h_2 \Pmax_2,\, h_2 < 1+h_1 \Pmax_1 \\
	(\Pmax_1,0), &\text{if } h_1 < 1,\, h_2 \ge 1+h_1 \Pmax_1 \\
	(0,0), & \text{otherwise}
	\end{cases}
\end{equation}
\end{theorem} \vspace{-.1in}
\begin{proof}
See Appendix \ref{app:GTWsum}.
\end{proof}
Note that the solution is such that as long as a user is not single-user decodable, it should be transmitting with maximum power.  Comparing this with the GGMAC-WT region found in \cite{tekin:ALLERTON06}, we note the same structure, namely, that secrecy is achievable for both users, as long as neither can be decoded by treating the other user as noise.

\section{Cooperative Jamming}
\label{sec:coopjam}
In \cite{tekin:ALLERTON06}, it was shown that for the GGMAC-WT, a user who ceases transmission to maximize the secrecy sum-rate may jam the eavesdropper and allow an increase in the remaining users' secrecy rate, or even allow a user to achieve a positive secrecy rate.  Similarly, in Theorem \ref{thm:GTWsum}, we see that when user 2 is single-user decodable, i.e. $h_2 \ge 1+h_1 \Pmax_1$, it must cease transmission in order to maximize sum rate.  We show that in this case, user 2 can similarly help user 1 increase its secret rate and/or achieve a positive secrecy rate.  This is achieved by letting user 2 transmit white Gaussian noise instead of actual codewords.  Since receiver 2 knows the transmitted codewords, it can subtract these from its received sequence to get a clear channel from user 1.  However, the eavesdropper, devoid of this \ital{side information}, sees more noise and the achievable secrecy capacity for user 1 (since we are reduced to the single user case, \cite{leung-hellman:gaussianwiretap} established that this is indeed the capacity) is increased as it is the difference of the capacity to user 1's channel to receiver 2 and its channel to the eavesdropper.  This is stated below:
\begin{theorem}
\label{thm:GTWjam}
The optimum power allocations for the cooperative jamming scheme described is
\begin{equation}
(\Popt_1,\Popt_2) = 
\begin{cases}
(\Pmax_1,\Pmax_2), & \text{if } h_1 < 1+h_2 \Pmax_2\\
(0,0), & \text{otherwise}
\end{cases}
\end{equation}
\end{theorem}
\begin{proof}
See Appendix \ref{app:GTWjam}.
\end{proof}
This can be interpreted as ``jam with maximum power if it is possible to change user 1's effective channel gain such that it is no longer single-user decodable".  If $h_2 < 1+h_1 \Pmax_1$, then user $2$ must be transmitting instead of jamming.

We can similarly consider a scheme for the BATW-WT.  Note that the achievable secret-sum rate is $0$ when $h(\ep_1)+h(\ep_2) \ge 1+h(\epW)$.  This implies that $\ep_k \ge \epW, \, k=1,2$. Let $\ep_1 \le \ep_2$.  Then, user 2 can randomly transmit bits drawn according to the binary distribution with $\prob{X_{2i}=1}=\onehalf$.  This is equivalent to randomly adding a bit to the eavesdropper's signal.  Hence the probability of error at the eavesdropper for user 2's codeword becomes $\onehalf$, and the eavesdropper cannot gain any information about user 1's transmitted codeword.  Receiver 2, however, knows the jamming sequence, which it can subtract from its received sequence.  Thus, user 1 can transmit to user 2 at a rate $1-h(\ep_1)$, which is its capacity.

\section{Numerical Results and Conclusions}
\begin{figure}[t]
\centering
\resizebox{2in}{!}{\input{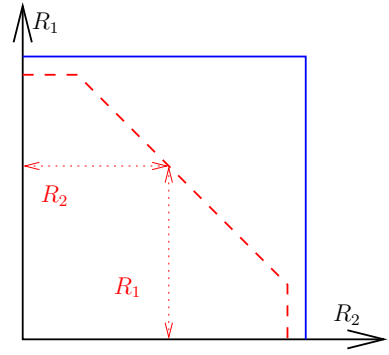}}
\caption{\small Example region for a TW-WT}
\label{fig:TWWTRegions}
\vspace{-.2in}
\end{figure}

\begin{figure}[t]
\centering
\includegraphics[width=3in,angle=0]{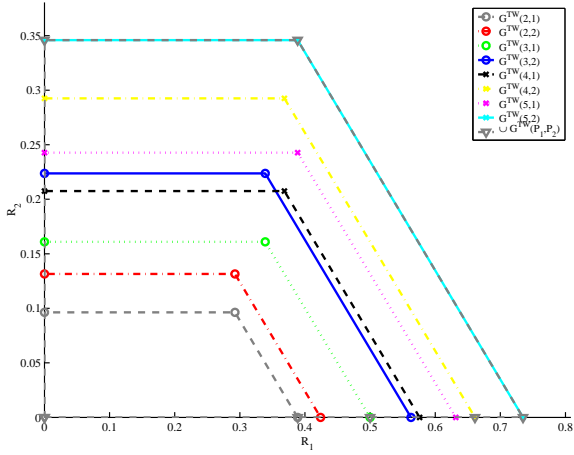}
\caption{\small GTW-WT achievable secrecy region when $\Pmax_1=5, \Pmax_2=2, h_1=.5, h_2=1.5$.}
\label{fig:GTWUreg}
\vspace{-.2in}
\end{figure}

\begin{figure}[!t]
\centering
\includegraphics[width=2.8in,angle=0]{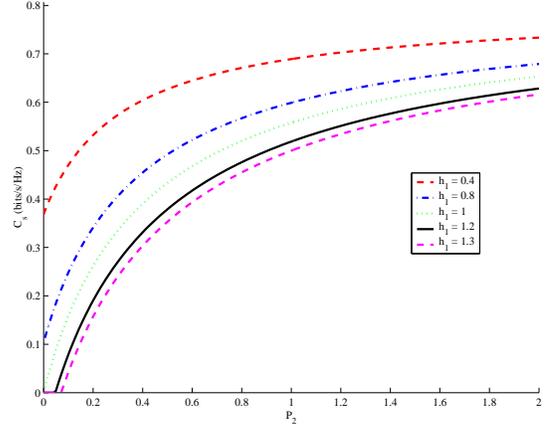}
\caption{\small GTW-WT cooperative jamming secrecy capacity as a function of $P_2$ with different $h_1$ for $\Pmax=2, \, h_2=4.2$}
\label{fig:GTWjam}
\vspace{-.2in}
\end{figure}

We now illustrate our results via numerical examples.  A typical region for the TW-WT channels considered is shown in Figure \ref{fig:TWWTRegions}.
Figure \ref{fig:GTWUreg} shows an achievable region as a function of the power allocations.  We can see that the optimum power allocation is given by Theorem \ref{thm:GTWsum}.  Finally, Figure \ref{fig:GTWjam} shows the secrecy capacity increase for user 1 as a function of user 2's jamming power.

Comparing the GTW-WT region to the achievable region for the GGMAC-WT, \cite{tekin:ALLERTON06}, we can see the enlargement of the rate region due to the two-way communications scenario.  Even though the signal received by the eavesdropper is the same, the two-way channel effectively provides a shared secret, namely the transmitters' knowledge on their own codewords, and hence enlarges the rate region.  In addition, unlike the GGMAC-WT, sum secret rate is not limited by the channel gains as the power limitations are relaxed for the GTW-WT.  In fact, we see that $\limtoinf{\Pmax} R_1+R_2 = g(\onehalf \Pmax)$.

An important thing to note is that in both channels, the achievability proof uses a scheme that requires the receivers to decode one of $\twon{R_k+\Rx_k}$ codewords.  Thus, the actual rate of communication is $R_k+\Rx_k$, although only a rate $R_k$ is ``secret" information.  We can utilize the extra codewords to communicate at an additional rate $\Rx_k$, although the secrecy of these messages is not guaranteed.  Thus, we may use the channel to its full capacity, but are limited in rate by how much of the communication can be kept secret as in \cite{leung-hellman:gaussianwiretap, tekin:IT06a}.

In conclusion, we see that two-way channels provide an extra advantage for wire-tap scenarios as the receivers, knowing their own transmitted codewords, gain an advantage over the eavesdropper that is not possible for multiple-access wire-tap channels.  As a result, a larger achievable region is found, and for the scenarios considered cooperative jamming proves to be even more useful as it does not hurt the transmitting user's rate as it does for the GMAC-WT.
\appendices
\newcommand{\Xc}{\mathfrak{X}}
\newcommand{\Xcx}{\tilde \Xc}
\renewcommand{\CW}{C_\Wch}
\newcommand{\XmS}{\Xm_\Sigma}

\section{Achievability Proofs}
\label{app:TWprf}
The proofs follow along the same line as the proof in \cite{tekin:ALLERTON06} for the achievability of the general GMAC-WT.
Let $\Pm \in \Ps$ and $\Rm \in \RsGTW$ for the GTW-WT, and let $\Rm \in \RsBATW$ for the BATW-WT.  Consider user $j=1$, and the following scheme (the other user does exactly the same):
\begin{IEEEenumerate}[
\setlength{\topsep}{0.0in}
\setlength{\parskip}{0in}
\setlength{\labelindent}{0.06in}
\setlength{\labelwidth}{0.05in}
\setlength{\labelsep}{3pt}]
\item	Generate $2$ codebooks $\Xc_1,\Xcx_1$.  $\Xc_1$ consists of $M_1$	codewords, and codebook $\Xcx_1$ has $\Mx_1$ codewords. The codebooks are generated such that
	\begin{IEEEenumerate}[
	\setlength{\labelindent}{0.06in}
	\setlength{\labelwidth}{0.05in}
	\setlength{\labelsep}{3pt}]
	\item For the GTW-WT, each component of the codes in $\Xc_1$ is drawn 
		$\isnormal{0,\lambda_1 P_1 -\varepsilon}$, and each component of $\Xcx_1$ is 
		drawn $\isnormal{0,(1-\lambda_1) P_1-\varepsilon}$ where $\varepsilon$ is an arbitrarily
		small number to ensure that the power constraints on the codewords are satisfied with
		high probability. 
	\item For the BATW-WT, codewords in $\Xc_1$ and $\Xcx_1$ are drawn uniformly 
		according to a binary distribution with $p=\onehalf$.
	\end{IEEEenumerate}
\item To transmit message $W_1 \in \Ws_1$, user $1$ finds the codeword corresponding to $W_1$ in $\Xc_1$ and also uniformly chooses a codeword from $\Xcx_1$. User 1 then adds (xor's for the binary case) these codewords and transmits the resulting codeword, $\Xm_1$, so that we are actually transmitting (uniformly) one of $M_1 \Mx_1$ codewords, and the rate of transmission is $R_1+\Rx_1$, where $\Rx_1=\ninv \log \Mx_1$.
\end{IEEEenumerate}
We choose the rates to satisfy
\begin{align}
R_k + \Rx_k &\le C_k, \quad k=1,2 \\
R_1 + R_2 &\le C_1 + C_2 - \CW \\
\Rx_1+\Rx_2 &= \CW
\end{align}

The first set of conditions for both channels guarantee that receiver $1$ can reliably decode the $\twon{R_2+\Rx_2}$ codewords from user $2$ since it knows its own transmitted codeword, $\Xm_1$, and can subtract (or xor for the binary case) this with its received sequence $\Ym_1$ to get a single-user channel from the other transmitter such that for the GTW-WT, it has $\Xm_2+\Nm_1$ and for the BATW-WT, it has $\Xm_2 \xor \Em_1$.  Then, the standard channel coding arguments, see \cite{cover-thomas:IT}, can be used to establish that rates of $C_k$ can be achieved for user $k$. Define 
\begin{equation}
\XmS=\case{\sqrt{h_1}\Xm_1 + \sqrt{h_2}\Xm_2}{\text{GTW-WT}}{\Xm_1 \xor \Xm_2}{\text{BATW-WT}}
\end{equation}
Then,
\begin{align}
\hspace{-.085in} H(\Wmsec|\Zm) &=H(\Wmsec,\Zm)-H(\Zm) \\
	&= H(\Wmsec,\XmS,\Zm)-H(\XmS|\Wmsec,\Zm)-H(\Zm) \\
	&=H(\Wmsec)+H(\Zm|\Wmsec,\XmS)-H(\Zm) \notag \\
	&\hspace{.5in} +H(\XmS|\Wmsec)-H(\XmS|\Wmsec,\Zm) \\
	&\label{eqn:achprf1}= H(\Wmsec) - I(\XmS;\Zm)+I(\XmS;\Zm|\Wmsec)
\end{align}
where the key observation is that the eavesdropper's information on $\Wm$ only depends on $\XmS$, i.e. $\Markov{\Wm}{\XmS}{\Zm}$, and hence $H(\Zm|\Wmsec,\XmS)=H(\Zm|\XmS)$.

Note that we have $I(\XmS;\Zm) \le n\CW$, from the capacity of the standard single user binary additive channel.  We can also write $I(\XmS;\Zm|\Wmsec)= H(\XmS|\Wmsec)-H(\XmS|\Wmsec,\Zm)$.  Since, given each pair of messages, $\Wmsec$, we uniformly send one of $\Mx_1 \Mx_2$ random sum-codewords, we have $H(\XmS|\Wmsec) = \log (\Mx_1 \Mx_2) = n \paren{\Rx_1 + \Rx_2} = n \CW$.  In addition, we have $H(\XmS|\Wmsec,\Zm) \le \e$, since given $\Wmsec$, we transmit one of only $n\CW$ codewords, and the eavesdropper can reliably decode these.  Thus, we also have $I(\XmS;\Zm|\Wmsec) \ge  n\CW-n\e$.  Using these in \eqref{eqn:achprf1}, we see that
\begin{equation}
H(\Wmsec|\Zm) \ge H(\Wmsec) - n\CW + n\CW -n\e =H(\Wmsec) -n\e
\end{equation}
\vspace{-.2in}
\newcommand{\rhodot}{\dot \rho}
\newcommand{\phidot}{\dot \phi}
\newcommand{\jc}{j^c}
\vspace{-.1in}
\section{Proof of Theorem \ref{thm:GTWsum}}
\label{app:GTWsum}
\vspace{-.05in}
The Lagrangian is given by,
\begin{equation}
\label{eqn:GTWLag}
\Lag(\Pm,\muv) = \rho(\Pm)
	-\ssum_{k=1}^2 \mu_{1k}P_k + \ssum_{k=1}^2 \mu_{2k}(P_k-\Pmax_k)
\end{equation}

Equating the derivative of the Lagrangian to zero,
\begin{equation}
\label{eqn:TWLagder}
\frac{\del \Lag(\Pmopt,\muv)}{\del \Popt_j} 
	= \rhodot_j(\Pmopt) -\mu_{1j} + \mu_{2j}  = 0 
\end{equation}
where 
\vspace{-.05in}
\begin{align}
\rhodot_j (\Pm) &\triangleq \frac{h_j-\Phi_j(\Pm)}{(1+P_1)(1+P_2)} 
\label{eqn:rhodotdef}\\
\Phi_j (\Pm) &\triangleq \frac{1+h_1 P_1 + h_2 P_2}{1+P_j}
\label{eqn:phidef}
\end{align}
It is easy to see that if $h_j > \Phi_j(\Pm)$, then $\mu_{1j}>0$, and we have $\Popt_j=\Pmax_j$.  If $h_j < \Phi_j(\Pm)$, then we similarly find that $\Popt_j=0$.  Finally, if $h_j = \Phi_j(\Pm)$, we can have $0<\Popt_j<\Pmax_j$.  However, such a user has $\rhodot_j(\Pmopt)=0$, so we can set $\Popt_j=0$ with no effect on the secrecy sum-rate.  Thus, we have $\Popt_j=\Pmax_j$ if $h_j < \phi_j(\Pm)$, and $\Popt_j=0$ if $h_j \ge \phi_j(\Pm)$.

Now consider user 1.  If $\Popt_1=0$, then $h_2 \ge h_1 \ge 1+h_2 \Popt_2$, but if $\Popt_2>0$, this implies that $h_2 > \frac{1+h_2 \Popt_2}{1+\Popt_2}$ and $\Popt_2=0$. This contradiction shows that if $\Popt_1=0$, then $\Popt_2=0$.  Assume $\Popt_1=\Pmax_1$, i.e. $h_1 < 1+ h_2 \Popt_2$.  If $h_2 > 1+h_1 \Pmax_1$, then $\Popt_2=0$.  If $h_2 < 1+h_1 \Pmax_1$, then $\Popt_2=\Pmax_2$.
\vspace{-.1in}
\section{Proof of Theorem \ref{thm:GTWjam}}
\label{app:GTWjam}
\vspace{-.05in}
We can write the problem formally as:
\begin{equation}
\max_{\Pm \in \Ps} \quad
		g\paren{P_1}-g\paren{\frac{h_1 P_1}{1+h_2 P_2}} 
	\equiv \min_{\Pm \in \Ps} \frac{\rho(\Pm)}{\phi_2 (P_2)}
\end{equation}
where $\rho$ is given in \eqref{eqn:rhodef} and $\phi_2(P_2) \triangleq \frac{1+h_2 P_2}{1+P_2}$.

The Lagrangian is given by 
\begin{equation}
\label{eqn:GTWjamLag}
\Lag(\Pm,\muv) = \frac{\rho(\Pm)}{\phi_2(P_2)}
	-\ssum_{k=1}^2 \mu_{1k}P_k + \ssum_{k=1}^2 \mu_{2k}(P_k-\Pmax_k) 
\end{equation}

Taking the derivative with respect to $\Popt_1,\Popt_2$, we get: \newline
\vspace{-\parskip}
\begin{align}
\frac{\rhodot_1(\Pmopt)}{\phi_2(\Popt_2)}-\mu_{11}+\mu_{21} =0 
	\label{eqn:GTWjamLagder1}\\
\frac{\rhodot_2(\Pm) \phi_2(\Popt_2)-\rho(\Pm) \phidot_2(\Popt_2)}{\phi_2^2(\Popt_2)} 
	-\mu_{12} + \mu_{22} =0
	\label{eqn:GTWjamLagder2} 
\end{align}
where $\rhodot$ is as given in \eqref{eqn:rhodotdef} and $\phidot_2(P) \triangleq \frac{h_2-\phi_2(P)}{1+P}$.

Consider user 1.  If we have $h_1 > 1+h_2 \Popt_2$, then we must have $\mu_{11}>0$ since the first and last terms in \eqref{eqn:GTWjamLagder1} would be positive, making $\Popt_1=0$.  Assume $\Popt_1 > 0 \Rightarrow \mu_{11}=0$.  If $h_1 < 1+h_2 \Popt_2$, then the first term is negative, and $\Popt_1=\Pmax_1$. If $h_1= 1+h_2 \Popt_2$, then the sum rate is zero, and we can set $\Popt_1=0$.  For user 2, it is very easy to see that since it only harms the jammer, the optimal jamming strategy should have $\Popt_2=\Pmax_2$, as long as user $1$ has $\Popt_1>0$.  This can also be seen by noting that $\rhodot_2(\Pm) \phi_2(\Popt_2)-\rho(\Pm) \phidot_2(\Popt_2) < 0$ for $\Popt_1>0$.

\vspace{-.05in}
\bibliographystyle{IEEEtran}
\bibliography{IEEEabrv_mod,etekin_shortfull}

\end{document}